\documentclass[letter]{aa} % for the letters 
\usepackage{natbib}
\bibpunct{(}{)}{;}{a}{}{,}
\usepackage{graphicx}
\usepackage{txfonts}

\begin{document}

   \title{
   Dust production scenarios in galaxies at $z\sim6$--$8.3$
  }

%   \subtitle{...}

\titlerunning{Dust production at $z\sim6$--$8.3$}

   \author{Aleksandra Le\'sniewska
          %\inst{1}
          \and
          Michał Jerzy Micha\l{}owski%\inst{1}
          }

   \institute{Astronomical Observatory Institute, Faculty of Physics, Adam Mickiewicz University, ul.~S\l{}oneczna 36, 60-286 Pozna\'n, Poland\\
              \email{aleksandra.lesniewska@amu.edu.pl}}

%   \date{Received XXX; accepted XXX}

% \abstract{}{}{}{}{} 
% 5 {} token are mandatory
 
  \abstract
  % context heading (optional)
  % {} leave it empty if necessary  
   {The mechanism of dust formation in galaxies at high redshift is still unknown.  Asymptotic giant branch (AGB) stars and explosions of supernovae (SNe) are possible dust producers, and non-stellar processes may substantially contribute to dust production, for example~grain growth in the interstellar medium (ISM).}
  % aims heading (mandatory)
   {Our  aim is to determine the contribution to dust production of AGB stars and SNe in nine galaxies at $z\sim6$--$8.3$, for which observations of dust have been recently attempted.
   }
  % methods heading (mandatory)
   {In order to determine the origin of the observed dust we have determined dust yields per AGB star and SN required to explain the total amounts of dust in these galaxies.}
  % results heading (mandatory)
   {We find that AGB stars were not able to produce the amounts of dust observed in the galaxies in our sample. In order to explain these dust masses, SNe would have to have maximum  efficiency and not destroy the dust which they formed.}
  % conclusions heading (optional), leave it empty if necessary 
   {Therefore, the observed amounts of dust in the  galaxies in the early universe were %likely 
    formed either by efficient supernovae or by a non-stellar mechanism, for instance the grain growth in the interstellar medium.}


   \maketitle
%
%________________________________________________________________

\section{Introduction}

Dust in the universe can absorb and re-emit up to 30$\%$ of starlight in the infrared \citep{Hauser2001}.
Hence, the formation of dust has become one of the most important topics in galaxy evolution. 

It is well known that there are two types of stellar sources that produce dust.
Asymptotic giant branch (AGB) stars, which are evolved stars with initial masses of 0.8--8.0 M$_\odot$, create dust in their cooling dense ejecta. This process is associated with an intense phase of mass loss due to stellar winds, up to 10$^{-4}$~M$_\odot$~yr~$^{-1}$ \citep{Bowen1991}.  They can dominate dust production as long as a burst of star formation took place at least 400\,Myr prior to observations
\citep{Dwek2007,Valiante2009,Dwek2011}.  One AGB star is able to produce 10$^{-5}$~--~10$^{-2}$ M$_\odot$ of  dust \citep{Morgan2003,Ferrarotti2006,Ventura2012,Nanni2013,Nanni2014,Schneider2014}. 

Supernovae (SNe) are another stellar source of dust. Dust formation takes place in the expanding ejecta a few hundred or thousand days after the explosion, and stellar progenitors have initial masses of 8--40 M$_\odot$. 

Observations of SN1987A in the Large Magellanic Cloud have revealed  that during such an event, up to 0.7 M$_\odot$ of dust could be created \citep{Matsuura2011}. Similarly, a large amount of dust, $\sim0.5 M_\odot$, has been reported for several supernovae \citep{Gall2014, Owen2015, Bevan2017, DeLooze2017, Temim2017, Rho2018, Chawner2019}. This means that in the early galaxies a large amount of dust can be formed by SNe \citep[e.g.][]{Gall2018}. However, it is possible that SNe destroy most of the dust they form by  reverse shock waves \citep{Temim2015,Bianchi2007,Cherchneff2010,Gall2011,Lakicevic2015}, but it is debated how much of the new and pre-existing dust is destroyed by a supernova as SN dust grains may be large and distributed in clumps \citep{Gall2014,Lau2015,Wesson2015,Bevan2016,Micelotta2016,Gall2018,Matsuura2019}.

Dust grains formed by AGB stars and SNe can act as seeds that grow in the ISM, and this process can lead to a significant increase in the total dust mass \citep{Draine1979,Dwek1980,Dwek2007, Draine2009}.
However, it is not clear if this process is efficient and quick enough, especially at high redshift. \citet{Ferrara2016}  show that it is too slow in the diffuse ISM, and probably prohibited in  molecular clouds because of icy mantles forming on the gas grains.

In this work we investigate a  sample of galaxies at $6<z<8.4$ (900 -- 600 Myr after the Big Bang) with the latest observational constraints on dust masses. Our aim is to test whether AGB stars or SNe are able to explain the observed amounts of dust in these galaxies, or whether dust accumulation must have also happened by a different (non-stellar) mechanism, for example grain growth in the ISM. We use a cosmological model with $H_0$ = 70 km s$^{-1}$ Mpc$^{-1}$, $\Omega_{\Lambda}$~=~0.7, and $\Omega_m$~=~0.3.

\section{Sample}

We have selected all galaxies at $z>6$ for which observations of dust continuum have been attempted, except  those for which a similar method has already been applied: quasars J1048+4637, J1148+5251, and J2054-0005 \citep{Michalowski2010}, and galaxies analysed in \citet{Michalowski2015}. We describe below the measurements used to estimate the dust and stellar masses needed for our analysis. 

\begin{figure*}[h]
\centering
   \includegraphics[width=18cm]{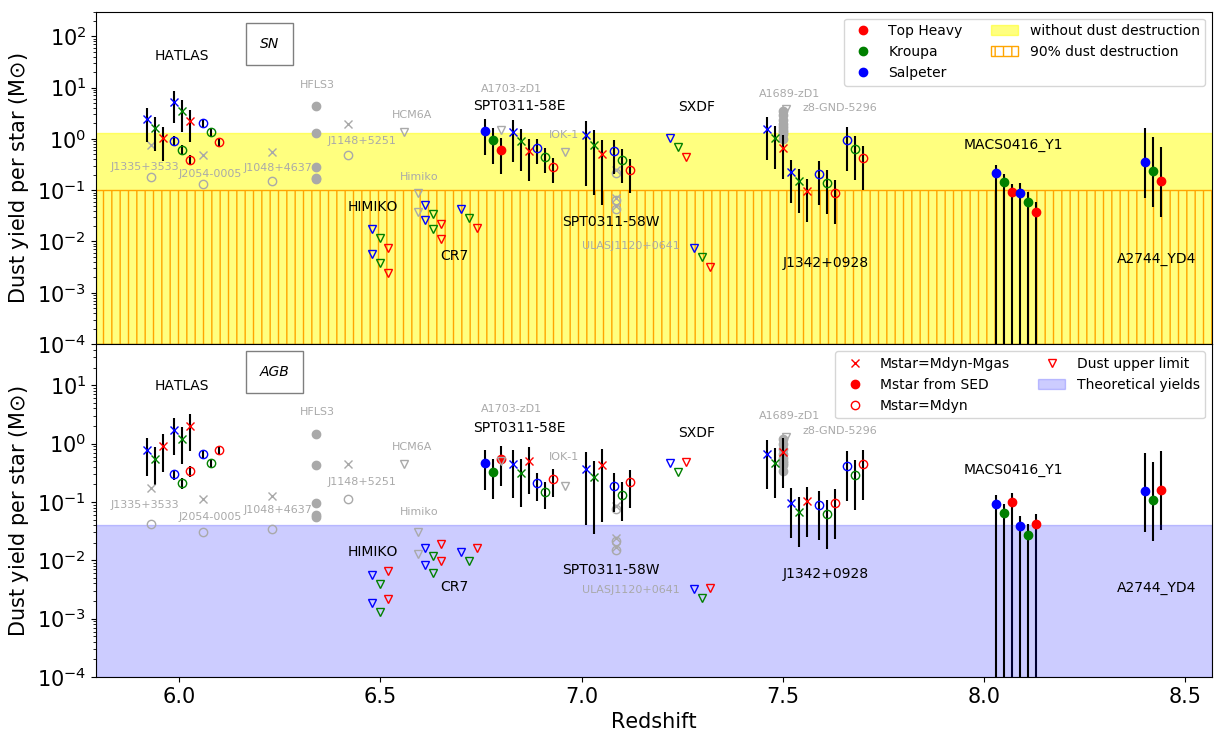}
     \caption{Dust yield per SN (top) and AGB star (bottom) required to explain the dust masses in galaxies in our sample. Three forms of the initial mass function are assumed: top-heavy (red), Kroupa (green), and Salpeter (blue). The stellar mass was determined in various ways: $M_{dyn}$-$M_{gas}$ (cross), $M_{dyn}$ (open circle), and  SED modelling (filled circle). In addition, galaxies with upper limits of  dust masses are indicated by down-pointing triangles. 
     Multiple points with the same shape and colour for a galaxy are for different dust or stellar mass estimates.
     Galaxies considered elsewhere \citep{Michalowski2010, Michalowski2015} are shown as grey symbols.
     For SNe  two regions have been defined: below the maximum theoretical dust yield without dust destruction of 1.3$M_\odot$ (yellow) and below the limit of 0.1$M_\odot$ including $\sim90$\% dust destruction  (vertical orange stripes).
     For AGB stars the  theoretically allowed dust yields are indicated (light blue).}
     \label{Fig1}
\end{figure*}

HATLAS\,J090045.4+004125 \citep[hereafter HATLAS;][]{HATLAS} is one of a few  submillimetre  galaxies above $z = 6$ (together with that reported by \citealt{Riechers2013}). HATLAS was selected using the {\it Herschel} Space Observatory within the {\it Herschel} Astrophysical Terahertz Large Area Survey \citep{Eales2010,Bourne2016,Valiante2016}. Emission lines for  $^{12}$CO(6-5) and $^{12}$CO(5-4)  and the 1\,mm continuum were detected by the Large Millimeter Telescope \citep{HATLAS}. 
The mass of molecular gas was determined from the  CO lines, whereas the dust mass was calculated in two ways: based on fitting photometric measurements to modified black-body function and via modelling using the MAGPHYS SED code (Table \ref{Tab1}; \citealt{HATLAS}). The dynamical mass was calculated using the isotropic virial estimator (equation 5 in \citealt{HATLAS}).

Discovered with the Subaru telescope at redshift z$\sim$5.95,  HIMIKO was at that moment the most luminous Ly$\alpha$ emitter \citep{Ouchi2009}. Recent observations of this object using ALMA in band 6 revealed the presence of three clumps. From the [CII] detections the size of HIMIKO and the velocity linewidth were measured \citep{HIMIKOmod}. The upper limit on the continuum emission of S158$\mu$m~<~27$~\mu$Jy was also reported. 
This is deeper than the previous ALMA data for this galaxy reported in \cite{Ouchi2013}. 

Cosmos Redshift 7 (CR7) located at z = 6.6 is the most luminous Ly$\alpha$ emitter, discovered using the Subaru Telescope \citep{Matthee,Sobral2015}. From the ALMA [CII] line detection and dust continuum upper limit,  \cite{CR7}  estimated the dynamical masses and dust masses found in two regions of this galaxy, Clump A and Clump C-2 (Table \ref{Tab1}).

SPT-S J031132-5823.4 (SPT0311–58) was discovered using the South Pole Telescope \citep{Mocanu2013}, and confirmed at a redshift of 6.9 \citep{Strandet2017}.  With high-resolution ALMA observations, \cite{SPT0311} detected continuum, [CII] 158 $\mu$m and [OIII] 88 $\mu$m lines of two components of SPT0311–58, named East and West, and determined their gas and dust masses. Only for SPT0311–58 East was the stellar mass  determined because no stellar emission was detected for the West component. 

SXDF-N1006-2 was discovered during a survey for Ly$\alpha$ Emitters (LAEs) by the Subaru Telescope \citep{Shibuya} and is located at z = 7.2. \cite{SXDF} used ALMA to detect the [OIII] 88 $\mu$m emission line and, assuming that the galaxy is a circular disk, determined its dynamical mass (Table \ref{Tab1}). The upper limit on the continuum flux at band 6 (1.33 mm) of $<0.042$~mJy was also obtained.

J1342+0928 is a quasar located at z = 7.54, first detected by \citet{Banados2018}. The detection of the [CII] 158 $\mu$m line emission  allowed \citet{J1342} to determine the dynamical mass of this galaxy based on the virial theorem, and the assumption that the [CII] emission comes from a rotating disk. From the detection of the 223.5 GHz continuum, the mass of dust in this galaxy was determined (Table \ref{Tab1}). The limit on the molecular gas mass was derived from the observations of the CO(3-2) line (Table \ref{Tab1}).

MACS0416$\_$Y1 is one of the brightest Lyman Break Galaxies (LBGs) at $z\sim8$ \citep{Infante2015,Laporte2015}. Using the detection of the [OIII] 88 $\mu$m line and the dust continuum, \citet{MACS} confirmed its redshift as 8.3118, and measured its dust mass (assuming a dust temperature of 40 and 50~K). From the optical emission the stellar mass was determined (Table \ref{Tab1}). In our analysis, the lower error bar for the stellar mass was modified  disregarding model solutions with ages lower than one million years (Y.~Tamura, private communication).

\begin{table*}
 \caption[]{\label{Tab1}List of physical properties of the  galaxies in our sample.}
\centering
\large
\begin{tabular}{lcccccc}
\hline
 & $z$  & M$_{dyn}$ & M$_{dust}$ & M$_{gas}$& M$_{stellar}$ & Ref\\
 &  & (10$^{10}$ M$_\odot$) & (10$^{7}$ M$_\odot$) & (10$^{11}$ M$_\odot$) & (10$^{9}$ M$_\odot$) & \\
\hline
\hline
HATLAS & 6.027 & 2.6 & 19$\pm$4 \quad 42$\pm$7 & 0.16$\pm$0.06 & --- & 1\\
\hline
HIMIKO & 6.595 & 1.168 $^{\dagger}$ & $<$0.16 $^{\dagger}$ & --- & 35$_{-26}^{+15}$  & 2, 3\\
\hline
CR7 & 6.604 & --- & --- & --- & 20 & 4\\
CR7 Clump A & 6.601 & 3.9$\pm$1.7 & $<$0.81 & --- & --- & 5\\
CR7 Clump C-2 & 6.598 & 2.4$\pm$1.9 & $<$0.81 & --- & --- & 5\\
\hline
SPT0311-58E & 6.9 & 7.7 $^{\dagger}$ & 40$\pm$20 & 0.4$\pm$0.2 & 35 $\pm$15 & 6\\
SPT0311-58W & 6.9 & 54.222 $^{\dagger}$ & 250$\pm$160 & 2.7$\pm$1.7 & --- & 6\\
\hline
SXDF & 7.2 & 5 & $<$0.29 $^{\dagger}$ & --- & 0.347$_{-0.166}^{+0.616}$ & 7\\
\hline
J1342+0928 & 7.54 & $<$15 \quad $<$3.2 & 24.5$\pm$18.5 & $<$0.12 & --- & 8 \\
\hline
MACS0416\_Y1 & 8.3118 & --- & 0.36$\pm$0.07 \quad 0.82$\pm$0.16 & --- & 4.8$_{-1.8}^{+6.8}$ \quad 5.1$_{-4.9}^{+7.1}$ & 9\\
\hline
A2744\_YD4 & 8.38 & --- & 0.55$_{-0.17}^{+1.96}$ & --- & 1.97$_{-0.66}^{+1.45}$ & 10 \\
\hline
\end{tabular}
\tablebib{$\dagger$ indicates the value determined in this work. (1)~\citet{HATLAS};
(2) \citet{HIMIKOclump}; (3) \citet{Ouchi2009};
(4) \citet{Sobral2015}; (5) \citet{CR7};
(6) \citet{SPT0311}; (7) \citet{SXDF}; (8) \citet{J1342};
(9) \citet{MACS}; (10) \citet{A2744}.}
\end{table*}

A2744$\_$YD4 at z = 8.38 is an LBG lensed by the Pandora Cluster. This object was observed in the Hubble Frontier Fields (HFF) by \cite{Abell2744}. The ALMA detection of the [OIII] 88 $\mu$m line allowed the confirmation of its redshift \citep{A2744}. Based on the dust continuum detection and the optical emission, \cite{A2744} estimated the mass of dust and stars in this galaxy (Table \ref{Tab1}).

\section{Method}\label{Method}
We calculated the dynamical masses for galaxies with emission line detections for which this was not reported.
This was the case for HIMIKO and SPT0311-58. Based on the detection of the [CII] emission line  \citep{HIMIKOclump,SPT0311}, the sizes of these galaxies were measured as 3.9$\pm$1.1~$\times$~1.7$\pm$1.1~kpc for HIMIKO, 2.2~kpc for SPT0311-58E, and 7.5~$\times$~2.0~kpc for SPT0311-58W. The [CII] linewidths were measured as 180$\pm$50~kms$^{-1}$ for HIMIKO, 500~kms$^{-1}$ for SPT0311-58E, and 1000~kms$^{-1}$ for SPT0311-58W. In order to calculate the dynamical masses we used eq.~5 in \citet{HATLAS}, based on the isotropic virial estimator. The results of these calculations are flagged in Table \ref{Tab1} with  dagger symbols  ($\dagger$).

Dust masses were not reported for HIMIKO and SXDF. We used the reported dust continuum upper limits to estimate the upper limits of the dust masses of these galaxies, assuming the dust temperature of 40 K and using eq.~(5) in \citet{Michalowski2009}, based on \citet{Taylor2005} and \citet{Hildebrand1983}.
A value of the emissivity index of $\beta=2$ was assumed. This gives conservatively low dust masses (see Fig.~3 in \citealt{Michalowski2010c}). In particular, if we adopted $\beta=1.5$, then we would obtain dust masses 2.7 times higher, which would make the stellar dust producers even less likely.
If we used this method for all galaxies, not only for those whose dust masses have not been calculated elsewhere, we would obtain masses a factor of $0.84\pm0.33$ higher. This would not change any of our conclusions.
The results of our adopted dust mass calculations are flagged in Table \ref{Tab1} with dagger symbols ($\dagger$).

Using the same methodology as presented by \citet{Michalowski2010b} and 
\citet{Michalowski2015},
we determined the amount of dust that one star would have to produce in order to explain the observed amount of dust in every galaxy. The number of dust-producing stars was estimated from the stellar masses in the studied galaxies. The stellar masses were estimated in three ways: (1) as equal to $M_{dyn}$ to obtain the maximum possible value, (2) as equal to $M_{dyn}$~-~$M_{gas}$, and (3) from SED modelling. The number of stars with masses between $M_0$ and $M_1$ can be calculated by the integration of the stellar initial mass function (IMF), according to the formula $N(M_0 - M_1) = M_{stellar} \int_{M_0}^{M_1} \xi(M) \mathrm{d}M / \int_{M_{min}}^{M_{max}} \xi(M) M \mathrm{d}M$, where $\xi(M)$ is an IMF parametrised as $M^{-\alpha}$. We assumed $M_{min}$~=~0.15, $M_{max}$ = 120 $M_\odot$, and three types of IMFs: the \citet{Salpeter1955} IMF with $\alpha$ = 2.35, the \citet{Kroupa2001} IMF with $\alpha$ = 1.3 in the mass range 0.15 - 0.5 $M_\odot$ and $\alpha$ = 2.3 in the mass range 0.5 - 120 $M_\odot$, and a top-heavy IMF with $\alpha$ = 1.5. The dust yield per star required to explain dust observed in a galaxy is then $M_{dust}/N(M_{0} - M_{1})$.

At the redshifts of the studied galaxies the time since the Big Bang was short, such that low-mass stars had not had time to leave the main sequence and start producing dust during the AGB phase. Based on a lifetime on the main sequence of 10$^{10}$~$\times$~[M/M$_\odot$]$^{-2.5}$ \citep{Kippenhahn1990}, we assumed that  at z < 7.0 only stars with masses between 3 to 8 M$_\odot$ would have had time to enter the AGB phase, whereas  at z $\ge$ 7.0 the range of 3.5--8~M$_\odot$ was assumed. For SNe we assumed that their progenitors had masses in the range of 8--40 M$_\odot$.

\section{Results and discussion}

Figure \ref{Fig1} shows the dust yield per star required to explain the dust mass for a given galaxy, using all possible combinations of dust and stellar masses.
Galaxies considered elsewhere \citep{Michalowski2010, Michalowski2015} are shown as grey symbols.
The top and bottom panels assume that the dust is produced by SNe and AGB stars, respectively. Some regions have been highlighted. In the top panel are shown the theoretical dust yield per SN without dust destruction ($<1.3\,M_\odot$; \citealt{Todini2001,Nozawa2003}) and with $\sim90$\% dust destruction ($<0.1\,M_\odot$; \citealt{Bianchi2007,Cherchneff2010,Gall2011,Lakicevic2015}). We note that it is difficult to constrain the fraction of newly formed dust destroyed by a reverse shock \citep{Micelotta2016}, so a weaker dust destruction is also possible.
In the bottom panel the maximum theoretical  dust yield per AGB star is shown (0.04$M_\odot$; \citealt{Morgan2003,Ferrarotti2006,Ventura2012,Nanni2013,Nanni2014,Schneider2014}). 

Depending on the data available, the stellar mass was obtained in various ways: as $M_{dyn}$-$M_{gas}$, as $M_{dyn}$, and from SED modelling. Only for one object, SPT0311-58E, was it  possible to determine the required dust yield per star with the three different assumptions on stellar mass, and they do not differ from each other by more than their uncertainties. 
For this galaxy the required dust yield per SN is about 0.1--1$M_\odot$. This is close to the maximum dust yield predicted by simulations and to the highest observed values. Therefore, it is in principle possible that dust in this galaxy was formed by SNe, but that would require weak dust destruction and very efficient production. We cannot accept a similar conclusion for  AGB stars. One AGB star would need to create between 0.1 to 1$M_\odot$ of dust to explain the dust mass in SPT0311-58E, which is significantly higher than the allowed values.

The dust yield per SN required to explain dust in the HATLAS galaxy is close to  1$M_\odot$, so if they are responsible for the bulk of the dust production, then they would need to be maximally efficient and not to destroy any dust. 
AGB stars cannot be responsible for the dust production in this galaxy because the required dust yield per star is more than 10 times higher than the theoretical value.

In the case of galaxies SPT0311-58W, J1342+0928, and A2744$\_$YD4 it is possible that SNe are responsible for the observed dust. Some dust destruction, but not a significant amount, by SNe would be allowed in these cases as the derived dust yields are above 0.1$M_\odot$. This is in line with the result of \citet{Gall2018} that dust in distant galaxies (including A2744$\_$YD4) was formed by SNe, which requires very little dust destruction. Again, the required dust yields per AGB star for these three galaxies are significantly above the theoretical limit, so AGB stars have not contributed substantially to the dust production in these galaxies.

For the remaining four galaxies in our sample the data are of insufficient quality to constrain the dust production mechanism.
HIMIKO, CR7, and SXDF only have  upper limits for the mass of dust,
so we cannot rule out either SNe or AGB stars as dust producers. We can only conclude that one SN in these galaxies could not produce more than 0.01$M_\odot$ of dust. This indicates that SNe in the early Universe are much less efficient than the maximum theoretical values, and casts doubts on the efficient SN dust production in the remaining galaxies in our sample \citep[see also][]{Hirashita2014}. Our limit is two times deeper than  that obtained by \citet{Hirashita2014}  because we used stellar masses that are two times higher. For the last galaxy, MACS0416$\_$Y1, we also cannot rule out the dust production mechanism  because the large uncertainty on the measurement of its stellar mass results in a derived dust yield consistent with zero.

In summary, AGB stars were not able to form dust in the majority of $z>6$ galaxies. Our results are conservative (leading to low required dust yields) because we include all stars that in principle could contribute to dust production. Stars with the masses close to our AGB lower limit ($3$ or $3.5\,M_\odot$) could have reached the AGB phase, but only if they were all born at the beginning of the galaxy evolution.

Supernovae would need to be maximally efficient and not to destroy the dust they formed \citep[as in][]{Hjorth2014,Gall2018}. 
One of the recent observations of SN 1987A indicates that dust can re-form and re-grow in post-shock region after being destroyed by the shock (\citealt{Matsuura2019}, but see \citealt{Biscaro2014}). This is consistent with high dust production efficiency of SNe. Similarly, the detection of SN dust in a 10\,000 year old SN remnant \citep{Lau2015,Chawner2019} indicates that dust is not efficiently destroyed by SNe.

It is  unclear, however,  whether {all} SNe can produce close to $1\,M_\odot$ of dust. If this is not the case, then some non-stellar mechanism is required, for example grain growth in the ISM. \citet{Asano2013}
found that dust mass accumulation is dominated by the grain growth in the ISM if the metallicity is higher than a threshold value of $0.3$ solar metallicity (or less if the star formation timescale is longer). 
This is likely for very dusty galaxies in our sample (HATLAS, SPT0311-58, J1342+0928), but more normal galaxies (MACS0416\_Y1, A2744\_YD4) may have lower metallicities. However, for A2744\_YD4 we derived very high required dust yields per star, so either its   metallicity is above this threshold or grain growth is  always efficient below it. 

We consider AGB stars and SNe separately, but in reality both contribute to dust production at the same time. However, this does not affect our conclusions because for the detected galaxies the required dust yields for AGB stars are approximately ten times higher than the allowed value. This means that AGB stars could  produce at most 10\% of the dust in these galaxies, and thus considering AGB stars and SNe at the same time would lead to revising down the required dust yield per SN by only  10\%.

The question remains of  the source of heavy elements building the dust grains in these galaxies. Theoretical work has shown that each SN can produce around $1\,M_\odot$ of heavy elements  \citep{Todini2001,Nozawa2003,Bianchi2007,Cherchneff2009}. This is close to the required dust yields per SN in our sample, so, as do \citet{Michalowski2010}, we conclude that SNe are efficient enough to produce heavy elements needed to build dust grains in these galaxies, even if they do not directly form most of the dust.

\section{Conclusions}

We determined the dust yield per AGB star and SN   required to explain the observed amount of dust in galaxies at redshift $6<z<8.4$. We obtained very high required dust yields per AGB stars, so  they were  not able to produce the majority of the dust in these galaxies. In most cases we accepted the hypothesis about the formation of dust by SNe, but they would need to be maximally efficient and not to destroy much dust. This suggests either that supernovae were efficient in producing dust in these galaxies or that a non-stellar mechanism was responsible for a significant fraction of dust mass accumulation, for example grain growth in the ISM. 

\begin{acknowledgements}
We thank the referee for helpful comments and  Yoichi Tamura for the clarification on the stellar mass measurements in \citet{MACS}.
M.J.M.~acknowledges the support of  the National Science Centre, Poland, through the POLONEZ grant 2015/19/P/ST9/04010;
this project has received funding from the European Union's Horizon 2020 Research and Innovation Programme under the Marie Sk{\l}odowska-Curie grant agreement No. 665778.
A.L. acknowledges the support from Adam Mickiewicz University in Poznań,
Faculty of Physics, through grant POWR.03.01.00-00-S157/17; this project has received funding from The National Centre for Research and Development. This research has made use of The Edward Wright Cosmology Calculator http://www.astro.ucla.edu/\textasciitilde wright/CosmoCalc.html \citep{Wright2006} and of NASA's Astrophysics Data System.
\end{acknowledgements}

%-------------------------------------------------------------------

\bibliographystyle{aa} % style aa.bst

\end{document}